\documentclass{PoS}

\title{\pschitt\ - A Python package for the modelling of atmoSpheric Showers and CHerenkov Imaging Terrestrial Telescopes
}

\ShortTitle{pschitt!}

\author{\speaker{Thomas Vuillaume}\\
        LAPP, Universit\'e Savoie Mont-Blanc, CNRS, France\\
        E-mail: \email{thomas.vuillaume@lapp.in2p3.fr}}

\author{Florian Gat\'e\\
       LAPP, Universit\'e Savoie Mont-Blanc, CNRS, France\\
       E-mail: \email{florian.gate@lapp.in2p3.fr}}

\author{Gilles Maurin\\
       LAPP, Universit\'e Savoie Mont-Blanc, CNRS, France\\
       E-mail: \email{gilles.maurin@lapp.in2p3.fr}}
       
\author{Jean Jacquemier\\
       LAPP, Universit\'e Savoie Mont-Blanc, CNRS, France\\
       E-mail: \email{jean.jacquemier@lapp.in2p3.fr}}
       
\author{Giovanni Lamanna\\
       LAPP, Universit\'e Savoie Mont-Blanc, CNRS, France\\
       E-mail: \email{giovanni.lamanna@lapp.in2p3.fr}}

\abstract{The simulation of atmospheric showers through Monte-Carlo processes as well as their projection into Imaging Atmospheric Cherenkov Telescopes (IACT) is long and very computing intensive. As these simulations are the most advanced ones from a physics point of view, they are not suited for simple tests.

Here we present a Python package developed in order to model atmospheric showers using different profiles and to image them with an array of IACT. This allows for first order studies of the influence of the primary photon energy and angular direction on the stereoscopic images. Its simplicity also makes it convenient for public dissemination and outreach as well as for teaching purposes.

This package has been developed to make the most out of the simplicity of Python but has also been optimised for fast calculations. It is developed in the framework of the ASTERICS H2020 project and as such is available as an open-source software.}

\FullConference{35th International Cosmic Ray Conference\\
		 10-20 July, 2017\\
		 Bexco, Busan, Korea}

\newcommand{\pschitt}{{\fontfamily{tgadventor}\selectfont ps\textsuperscript{2}chitt!}}

\usepackage{gensymb}
\usepackage{graphicx}
\usepackage{subcaption}
\usepackage{amsmath}

\begin{document}

\section{Introduction}

Monte-Carlo simulations of electromagnetic showers is a central part of Imaging Atmospheric Cherenkov Telescope. They are used to better understand the physics at play in particle showers, to test and develop the reconstruction algorithms and to simulate and produce the instruments response function. However, these simulations are computationally intense and require important resources and time and are not always suited for simple tests.

Nonetheless, there are situations where simple tests are sufficient.
That is why we developed an algorithm to model atmospheric showers and project their images in Imaging Atmospheric Cherenkov Telescopes. This algorithm has been developed in Python for modularity and quick developments but the most heavy calculations have been optimised to limit computation times and resources.

\pschitt\ potential applications are various including but not limited to creating fake images stereoscopically coherent, testing arrays configuration or public dissemination.

\section{Modelling of Cherenkov showers}

In \pschitt, objects are modelled as a set of point-like particles positioned in space. To model atmospheric showers, the user can choose in a variety of distributions or define its own. Implemented distributions relevant to Cherenkov shower modelling include ellipsoids with uniform, random ellipsoids or Gaussian particle distributions. Specific hadronic or electromagnetic showers distributions as described in sections \ref{sec:hadronic} and \ref{sec:electromagnetic}.

\subsection{Hadronic showers\label{sec:hadronic}}

Hadron-induced atmospheric showers can be divided in three components: nuclear, hadronic and electromagnetic. Since The electromagnetic component account for roughly 90\% of the total calorimetric energy lost by the shower in the atmosphere, only this component is modelled. The longitudinal and lateral shower developments can be described using only three parameters: the energy of the primary particle ($E$), the first interaction depth ($X_1$) and the mass of the primary ($A$).  

The longitudinal profile (the number of particles in the shower front), can be described by the Greisen-Iljina-Linsley parametrisation (GIL) \cite{GIL}, introducing the concept of the age $s$ of the shower. It is derived from the Greisen relations \cite{greisen} and describes the number of electrons and positrons, $N$, for nucleus-initiated showers \cite{nucleus} as a function of the mass of the primary A, its energy $E_p$:

\begin{equation}
N(E_p,A,t) = \frac{E_p}{E_l} \times \exp(t(1-2\ln(s))-t_\text{max})
\end{equation}

With:

\begin{equation}
t=\frac{X-X_1}{X_0}\; , \;\;\;\;\;\; t_\text{max} = a+b\left ( \ln \frac{E_p}{E_c} -\ln A  \right ) \;\;\;\;\;\; \text{and} \;\;\;\;\;\; s=\frac{2t}{t+t_\text{max}}
\end{equation}

Where $X_1$ is the first interaction depth (in g/cm$^2$), $X_0$ = 36.7 g/cm$^2$ is the radiation length via bremsstrahlung of electrons in the air, $E_l = 1450$ MeV is a normalisation factor. The Greisen formula gives the critical energy $E_c = 81$ MeV at which the ionisation and bremsstrahlung rates become equal. $b=0.76$ is the value of the elongation rate obtained from adjusted data (see \cite{GIL}) and $a=1.7$ is an offset parameter. $X$ is the atmospheric depth in g/cm2 measured from $X$ = 0 g/cm$^2$ so that if $X$ < $X_1$, then $s$ < 0 and $N$ is undefined, interpreted as no secondary particles are yet created before the first interaction depth.

For a 3D description, the particle density $\rho$ as a function of the shower axis $r$ and $X_\text{max} - X_1$ is also taken into account. We use the lateral distribution proposed by Nishimura, Kamata and Greisen \cite{NKG}:

\begin{equation}
\rho(r) = N_s \times C(s) \left ( \frac{r}{r_m} \right )^{s-2} \left ( 1+ \frac{r}{r_m} \right )^{s-4.5}
\end{equation}

with:

\begin{equation}
C(s) = \frac{\Gamma(4.5 -s)}{2 \pi r^2_m \Gamma(s) \Gamma(4.5-2s)}
\end{equation}

Where $r$ is the distance to the shower axis in meter, s is the shower age parameter, $r_m$ is the Moli\`ere radius, $N_s$ is the shower size, which is the integrated number of particles in the shower. The Moli\`ere radius depends on the characteristics of the material (the air in this case) and is the radius around the shower axis wherein 90\% of the secondary particles are located.

The first interaction point $X_1$ depends of the interaction cross section of the primary hadron with the air. The probability can be written as:

\begin{equation}
\frac{\text{d}P}{\text{d}X_1} = \frac{1}{\lambda} \times \exp \left ( \frac{-X_1}{\lambda} \right ) \;\;\;\;\;\; \text{with} \;\;\;\;\;\; \lambda = \frac{\left < M_\text{air} \right >}{\sigma}
\end{equation}

Where $\left < M_\text{air} \right >$ is the atomic mass of the air (in g) and $\sigma$ is the interaction cross section of the primary hadron with the air (in cm$^2$)

\subsection{Electromagnetic showers \label{sec:electromagnetic}}

The longitudinal profile, (i.e. the evolution of the number of secondary particle in the shower) $N(t)$ is proportional to the calorimetric energy loss of the shower in the atmosphere. The energy loss can be expressed as follow in radiation length unit $t=X/X_0$ \cite{LonEM}:

\begin{displaymath}
N(t) \propto \frac{\partial E (t)}{\partial t} = E_0 \beta  \frac{(\beta t)^{\alpha-1}  e^{-\beta t}}{\Gamma(\alpha)} 
\end{displaymath}

Where $E_0$ is the energy of the primary photon, $\alpha$ and $\beta$ are phenomenological parameters. A fit to CORSIKA \cite{corsika} simulations gives:

\begin{displaymath}
\alpha/\beta=2.16+0.99 \ln(y) \;\;\; ; \;\;\; 1/\beta = 1.53+0.01 \ln(y)
\end{displaymath}

With $y=E/E_\text{c}$, $E_\text{c} = 86$ MeV in air and is the electrons critical energy, i.e., the energy at which the ionisation loss rate becomes equal to radiation losses. $\Gamma(\alpha)$ denotes the gamma function:

\begin{displaymath}
\Gamma(\alpha) = \int_0^\infty e^{-X} X^{\alpha-1} dX
\end{displaymath}

The lateral distribution is drawn from an exponential decrease model as a function of the distance from the shower axis, scaled by the Moli\`ere radius, at which 90\% of the energy is contained. 

\begin{displaymath}
R_M = \frac{X_0 E_s}{\rho E_c} \;\;\; \; \;\;\; E_s = m_e c^2\sqrt{4\pi \alpha}
\end{displaymath}

$m_e$ is the mass of the electron, $\alpha$ is the electromagnetic coupling constant and $\rho$ is the density of the atmosphere.

\subsection{Coordinates conversion}

For both electromagnetic and hadronic showers, the longitudinal profiles are describes in terms of crossed atmospheric depth $X$. This allows to draw shower profiles universally, in dependently from from the arrival direction of the primary particle. The signal observed at the ground level depends on the geometric distance of the Cherenkov emission to the location of the simulated experiment. Thus, one needs to convert the $X$ into geometric distance $\ell$, calculated numerically by integrating the atmosphere density along the shower axis:

\begin{equation}
X(\ell) =  \int_{\ell}^{\infty} \rho(z(\ell')) \, \mathrm{d}\ell'
\end{equation}

However, for computing time purposes, we use the following relation:

\begin{equation}
X(\ell) =  X_\text{v} (z(\ell)) / \cos \theta
\end{equation}

$X_\text{v} (z)$ represents the vertical atmospheric depth; it is known as the Linsley's parametrisation when considering the US Standard atmospheric model \cite{Usstd} and provides the integrated atmospheric depth traversed vertically from "infinity" (i.e. where $\rho$ is negligible, before entering the atmosphere) to altitude~$z$.

\section{Modelling of imaging telescopes}

Each telescopes is described by a position in space representing its mirror centre and an observing direction. Mirrors are modelled as perfect parabolic mirrors with a parametrised focal length. A camera is associated to each telescope and placed at a parametrised distance above the mirror. A standard modelling would place them at the focal distance. The cameras are described by a list of pixels positions.

Showers in the atmosphere are then imaged by the array of telescopes via a geometric projection in the mirrors focal plane. The camera mask is then applied to get a digitised image. Each particle is imaged and counted as one photon in the corresponding pixel. Electronic noise can be modelled by the adding of a Poisson distribution. 

\section{Example of atmospheric shower modelling \label{sec:example_shower3d}}

\begin{figure}
\centering
\includegraphics[width=0.75\textwidth]{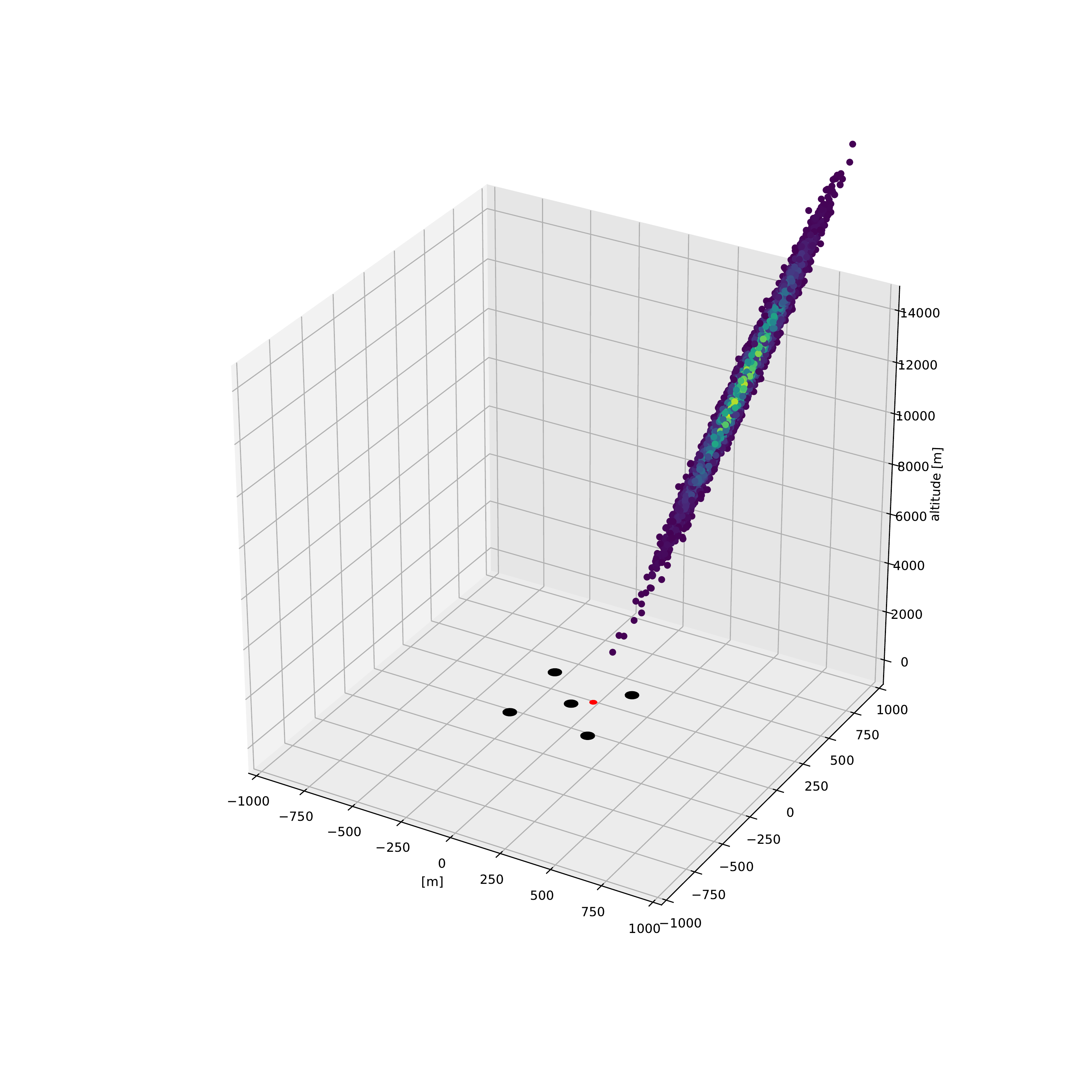}
\caption{3D representation of a shower and telescopes. Description of the shower is done in the text. Shower's impact parameter at [80,60,0] is represented by the red dot. The colours give the density of particles. The telescopes represented by the black dots are positioned at: [200,200,0], [-200, 200, 0], [-200, -200, 0], [200, -200, 0] and [0,0,0]. Their pointing direction is common (altitude=81\degree, azimuth=89\degree). \label{fig:3dshower}}
\end{figure}

Here we present an example of atmospheric shower modelling and imaging realised with \pschitt. From a centre placed at [0,0,8000], $10^4$ particles are placed following normal distributions with a standard deviation of 20 on the radial axis and of 800 on the z axis. The shower is then shifted to have an impact parameter at [80,60,0] and tilted to have a main axis direction of direction (altitude = 80\degree, azimuth = 90\degree).
Five telescopes are then placed on the ground at an altitude of $z=0$.
This is represented in figure \ref{fig:3dshower}.

\section{Resulting images}

The atmospheric shower given in section \ref{sec:example_shower3d} can then be imaged by the array of telescopes. The telescopes all point in the same direction (altitude=81\degree, azimuth=89\degree) and have a focal length of 16. The square cameras of side length equal 1 are placed in their focal plan. They have 2500 pixels here.

Each particle of the shower is projected in the mirrors focal plan following geometrical optics laws and the result is digitised to obtain the cameras images. An electronic noise following of Poisson distribution of expected occurrence equal to 6000 is added.

\begin{figure}[h!]
\centering
\begin{subfigure}{.32\textwidth}
  \centering
  \includegraphics[height=1\linewidth,trim={0 0 3cm 0},clip]{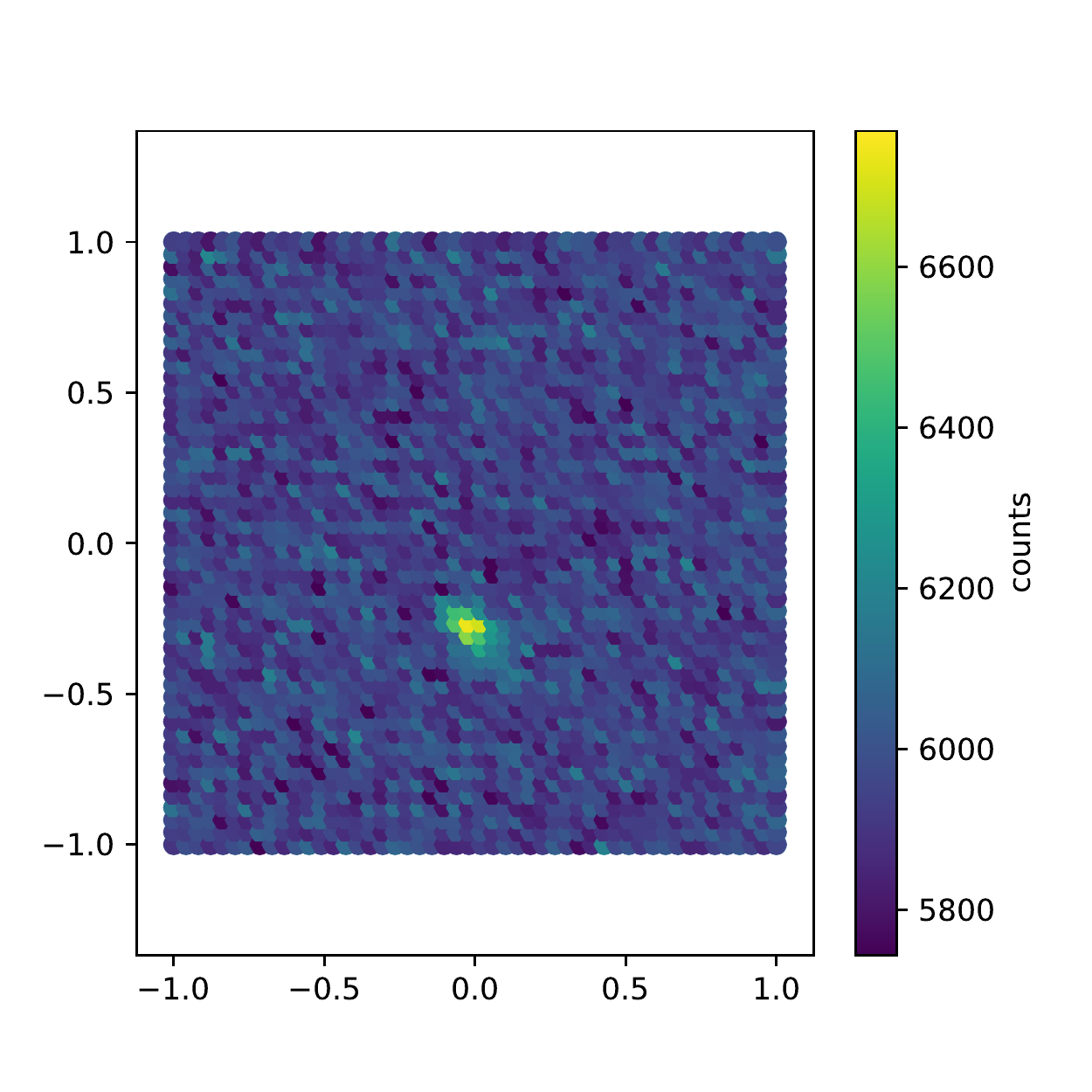}
  % \caption{A subfigure}
\end{subfigure}%
\begin{subfigure}{.32\textwidth}
  \centering
  \includegraphics[height=1\linewidth, trim={0 0 3cm 0},clip]{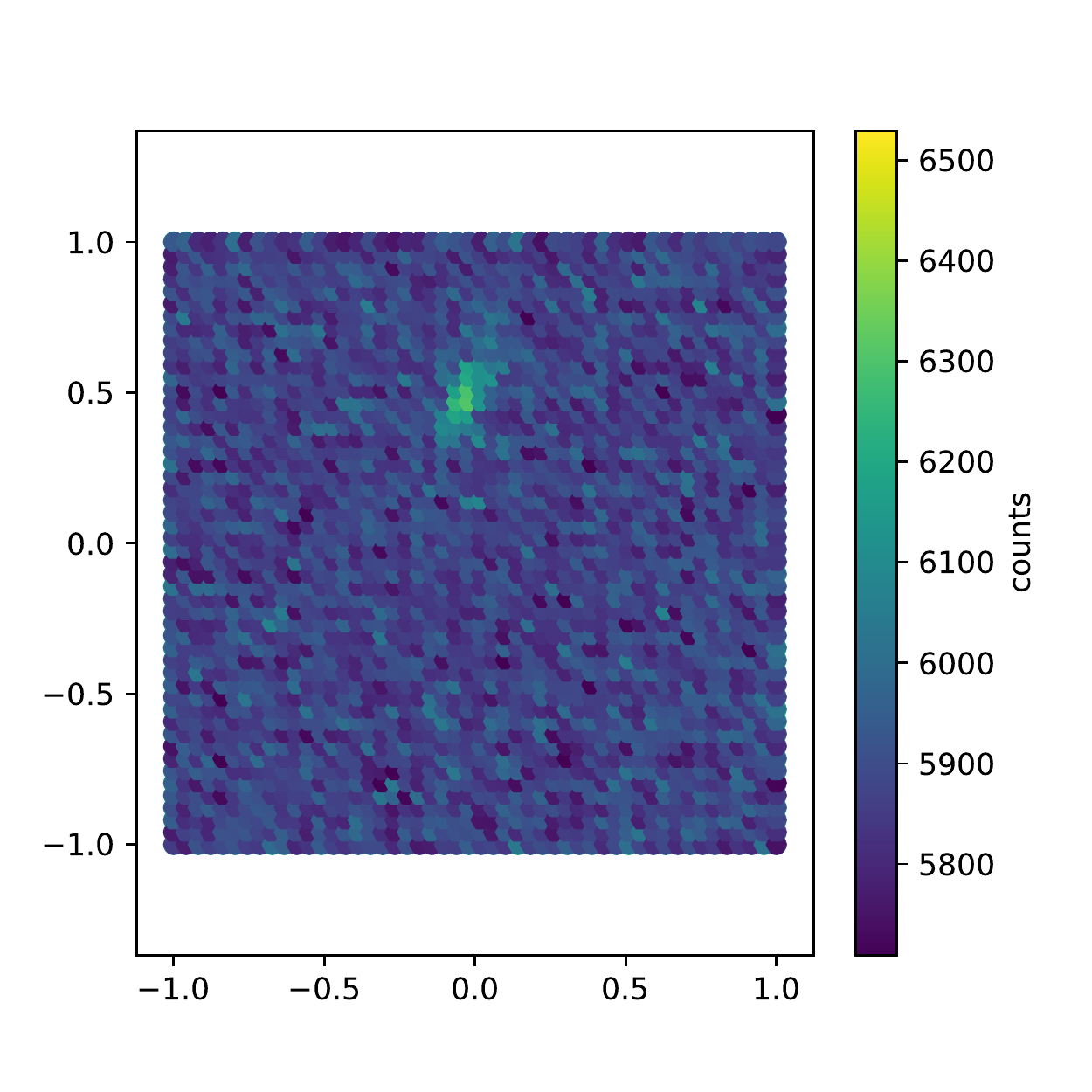}
  % \caption{A subfigure}
\end{subfigure}
\begin{subfigure}{.32\textwidth}
  \centering
  \includegraphics[height=1\linewidth, trim={0 0 3cm 0},clip]{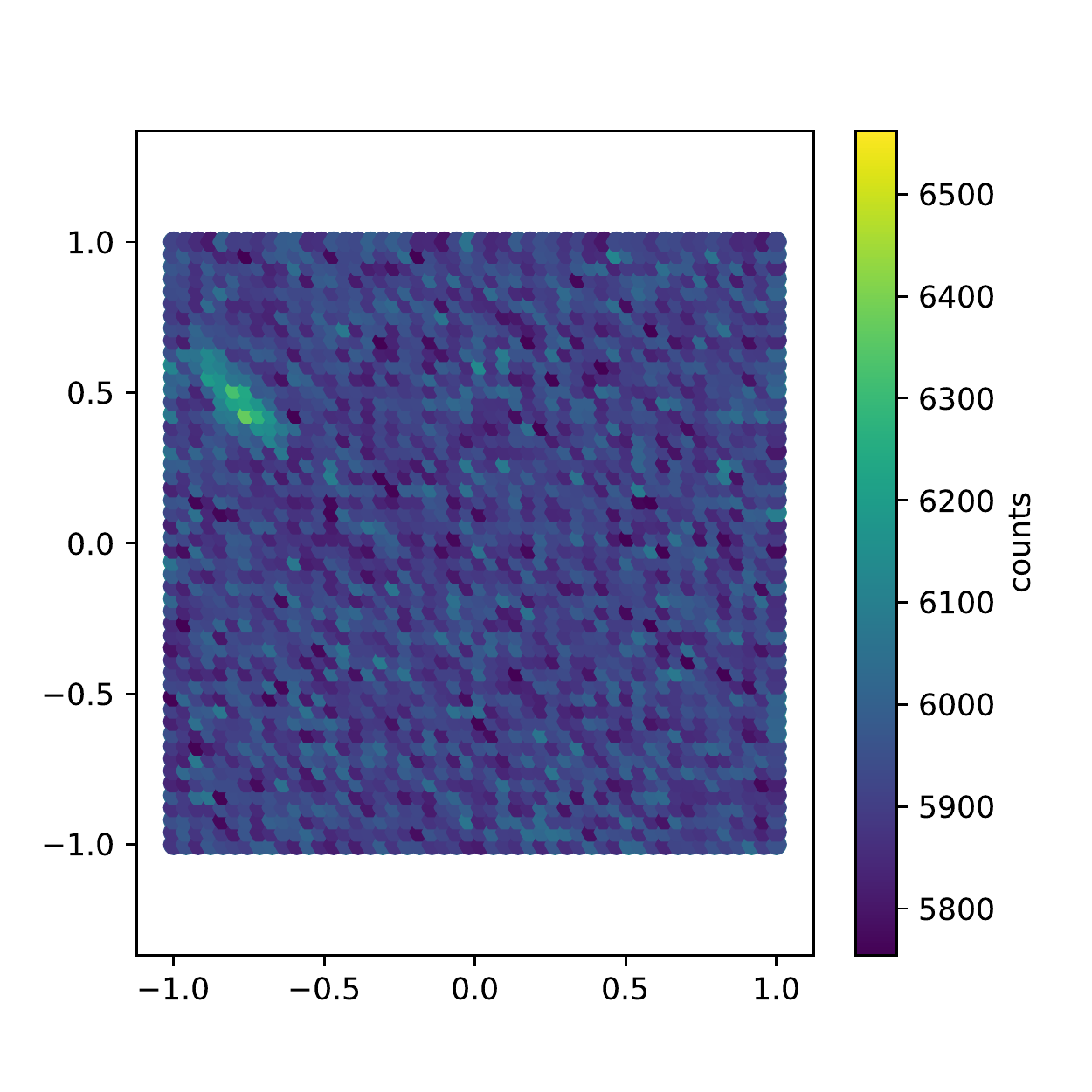}
  % \caption{A subfigure}
\end{subfigure}
\begin{subfigure}{.32\textwidth}
  \centering
  \includegraphics[height=1\linewidth,  trim={0 0 3cm 0},clip]{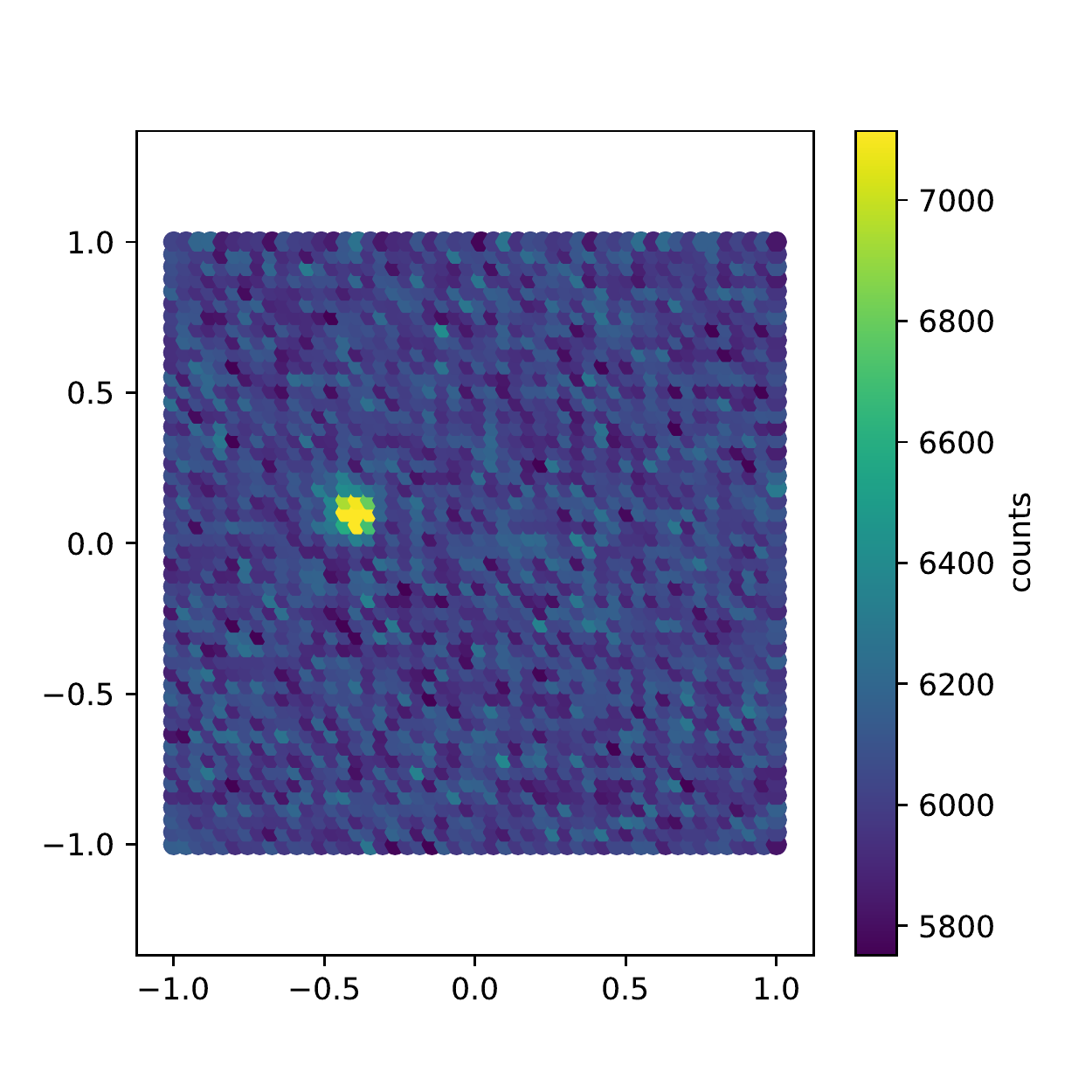}
  % \caption{A subfigure}
\end{subfigure}
\begin{subfigure}{.32\textwidth}
  \centering
  \includegraphics[height=1\linewidth]{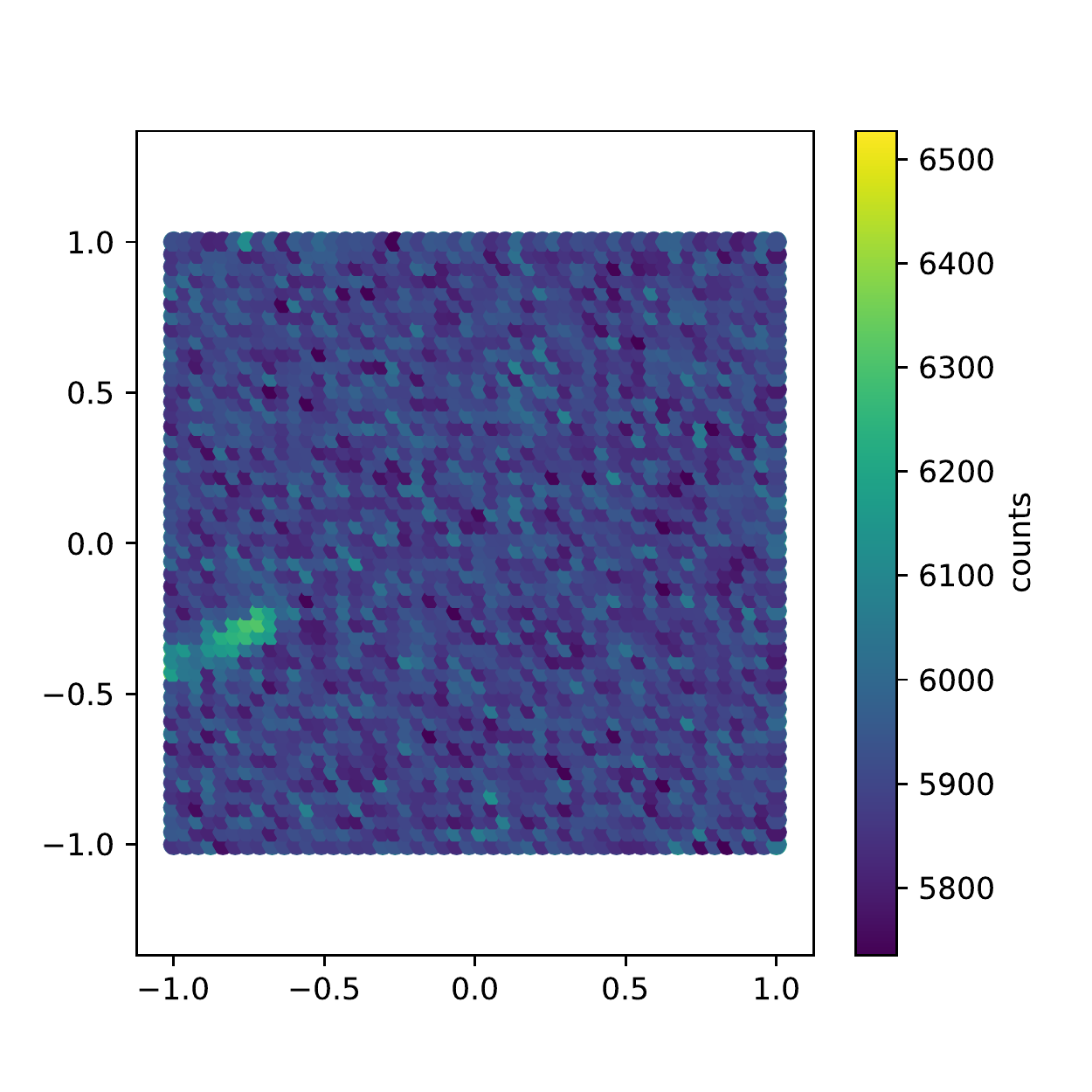}
  % \caption{A subfigure}
\end{subfigure}
\caption{Shower camera images of four telescopes of the array (see figure \ref{fig:3dshower} for the used configuration). Noise has been added. \label{fig:cameras}}
\end{figure}

Two examples of resulting images are given in figure \ref{fig:cameras}.

\section{Stereoscopic reconstruction}

A calculation of the Hillas parameters \cite{hillas} on the generated images, as well as a simple geometric reconstruction of the shower direction based on the intersection of the ellipses directions has been implemented. This can be used to do a reconstruction of the shower axis direction (equivalent to the sky source direction) and of the impact parameter of the shower.

\begin{figure}[h!]
\centering
\begin{subfigure}{.49\textwidth}
  \centering
  \includegraphics[width=0.8\linewidth,trim={0 0 3cm 0},clip]{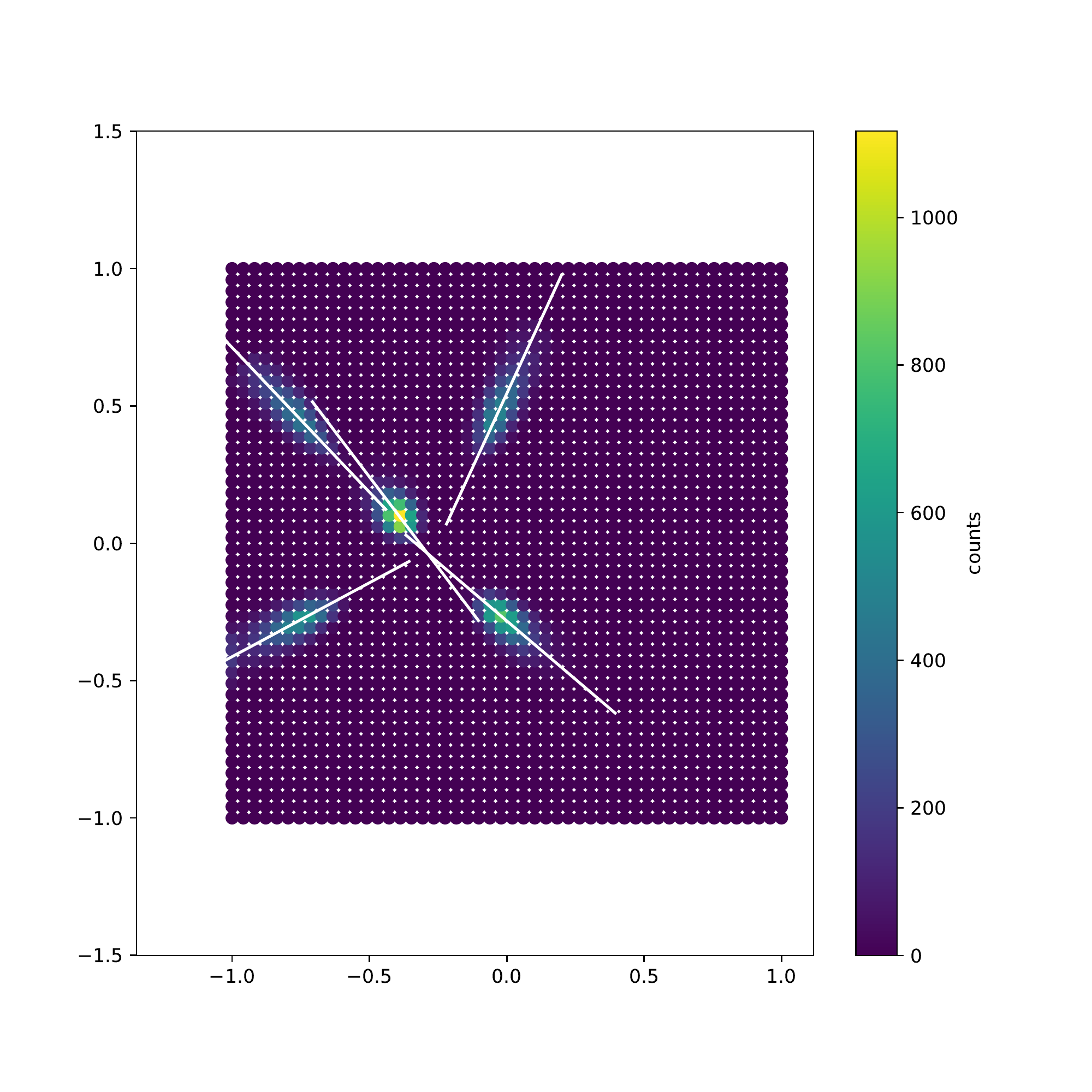}
  % \caption{A subfigure}
\end{subfigure}%
\begin{subfigure}{.49\textwidth}
  \centering
  \includegraphics[width=0.8\linewidth]{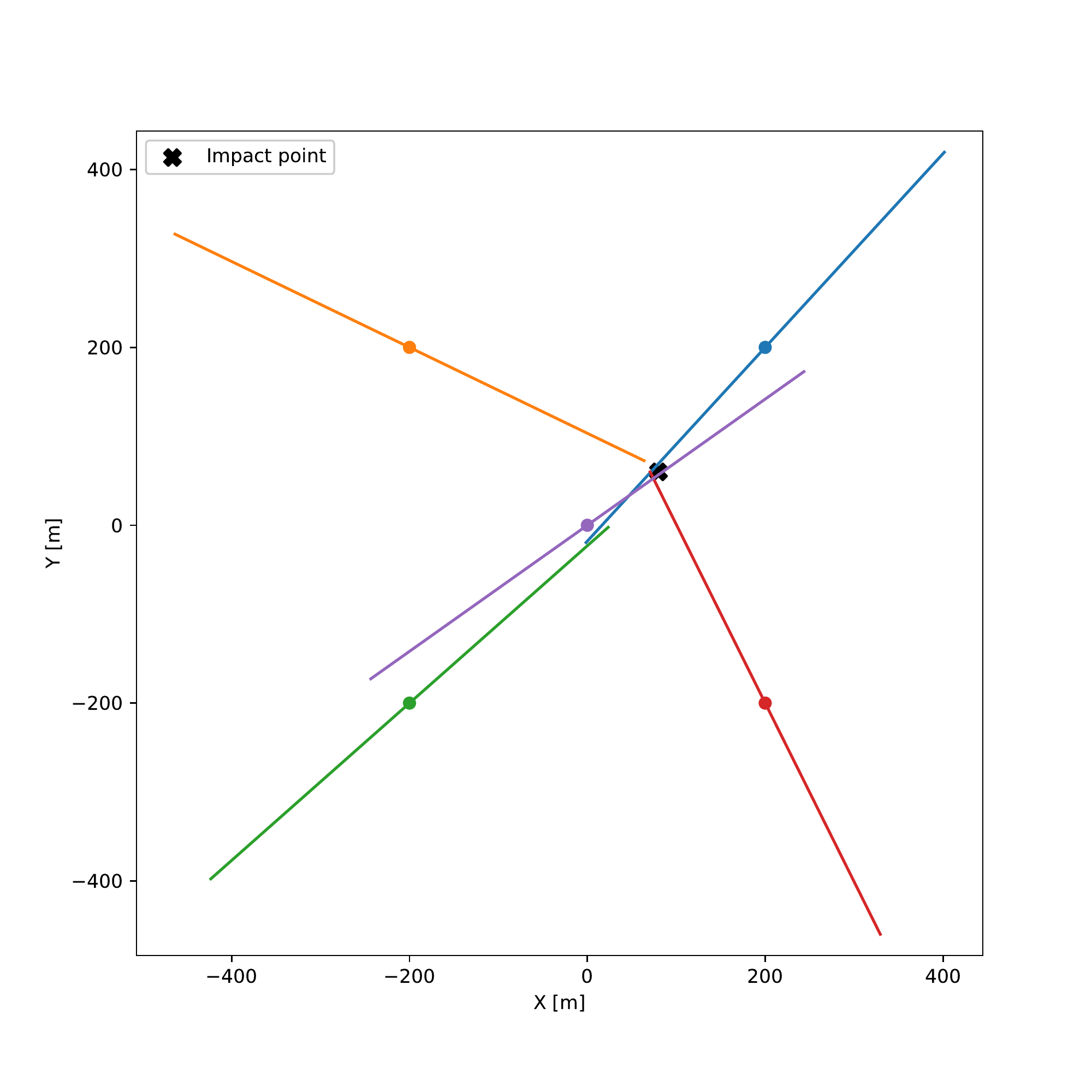}
  % \caption{A subfigure}
\end{subfigure}
\caption{Example of geometric reconstruction based on configuration given in figure \ref{fig:3dshower}. On the left, all the denoised camera images have been stacked to retrieve the direction of the shower in the sky. On the right, the ellipses directions are given in the site frame to retrieve the impact parameter. Here the reconstruction gives an impact parameter at [73.4, 55.6, 0.0] corresponding to an absolute error of 7.96 units. \label{fig:hillas}}
\end{figure}

An example of reconstruction based on previous configuration (see figure \ref{fig:3dshower}) is presented in figure \ref{fig:hillas}.

\section{Conclusion and acknowledgement}

We introduced a new Python package for the modelling of atmospheric showers and their imaging with and array of ground based Cherenkov telescopes. This software does not intend to replace Monte-Carlo but to substitute them for quick, simple and flexible modelling of showers. Arrays configuration (including telescopes positions and pointing) can quickly be tested and resulting images produced. \pschitt\ may already be downloaded \cite{pschitt} and used to generate fake images stereoscopically coherent, testing arrays configurations as well as for teaching and public dissemination and outreach.

We acknowledge support from the ASTERICS project supported by the European Commission Framework Programme Horizon 2020. Research and Innovation action under grant agreement n. 653477

\end{document}